\documentclass[pra,twocolumn]{revtex4-2}

\usepackage[pdftex]{graphicx}
\usepackage{siunitx}
\usepackage{amsfonts}
\usepackage{amsmath}
\usepackage[hidelinks]{hyperref}
\usepackage{csquotes}

\newlength{\onecolfig}
\newlength{\twocolfig}
\newcommand{\eqnref}[1]{eq.~(\ref{#1})}
\newcommand{\citeref}[1]{ref.~\cite{#1}}
\newcommand{\figref}[1]{fig.~\ref{#1}}
\newcommand{\secref}[1]{section~\ref{#1}}
\newcommand{\ish}{\mbox{$\sim$}\,}
\newcommand{\ltish}{\protect\raisebox{-0.4ex}{$\,\stackrel{<}{\scriptstyle\sim}\,$}}

\newcommand{\RF}{RF}
\newcommand{\DC}{DC}
\newcommand{\hessian}[1]{\mathbf{H}_{#1}}
\newcommand{\FPGA}{FPGA}
\newcommand{\kvec}{\vec{k}}
\newcommand{\caplus}{\textsuperscript{43}Ca\textsuperscript{+}}
\newcommand{\srplus}{\textsuperscript{88}Sr\textsuperscript{+}}
\newcommand{\charge}{Z e}  
\newcommand{\wRF}{\Omega_{\textrm{rf}}}
\newcommand{\wAM}{\omega_{\textrm{am}}}
\newcommand{\wm}{\omega_{\textrm{0}}}
\newcommand{\dcpot}{\Phi_{\textrm{dc}}}
\newcommand{\rfpot}{\Phi_{\textrm{rf}}}
\newcommand{\amat}{\mathcal{A}}
\newcommand{\qmat}{\mathcal{Q}}
\newcommand{\ee}{\mathrm{e}}
\newcommand{\ii}{\mathrm{i}}
\newcommand{\h}{h}
\newcommand{\hred}{\tilde{h}}
\newcommand{\prindex}{n}
\newcommand{\defeq}{:=}
\newcommand{\re}[1]{\operatorname{re}\left(#1\right)}
\newcommand{\im}[1]{\operatorname{im}\left(#1\right)}
\newcommand{\diff}[2]{\frac{\operatorname{d}\!{#1}}{\operatorname{d}\!{#2}}}
\newcommand{\difftwo}[2]{\frac{\operatorname{d}^2\!{#1}}{\operatorname{d}\!{#2}^2}}

\renewcommand{\vec}[1]{\boldsymbol{#1}}

\begin{document}

\title{Micromotion minimisation by synchronous detection of parametrically excited motion}

\author{D.~P.~Nadlinger}
\email{david.nadlinger@physics.ox.ac.uk}
\author{P.~Drmota}
\author{D.~Main}
\author{B.~C.~Nichol}
\author{G.~Araneda}
\author{R.~Srinivas}
\author{L.~J.~Stephenson}
\author{C.~J.~Ballance}
\author{D.~M.~Lucas}
\affiliation{Department of Physics, University of Oxford, Clarendon Laboratory, Parks Road, Oxford OX1 3PU, U.K.}

\begin{abstract}
	Precise control of charged particles in radio-frequency (Paul) traps requires minimising excess micromotion induced by stray electric fields.
	We present a method to detect and compensate such fields through amplitude modulation of the radio-frequency trapping field.
	Modulation at frequencies close to the motional modes of the trapped particle excites coherent motion whose amplitude linearly depends on the stray field.
	In trapped-ion experiments, this motion can be detected by recording the arrival times of photons scattered during laser cooling.
	Only a single laser beam is required to resolve fields in multiple directions.
	We demonstrate this method using a \srplus{} ion in a surface electrode trap, achieving a sensitivity of $\SI{0.1}{V m^{-1} / \sqrt{\hertz}}$ and a minimal uncertainty of \SI{0.015}{\volt \metre^{-1}}.
\end{abstract}

\maketitle

\section{Introduction}

Paul traps confine charged particles in all three spatial dimensions through a superposition of static and oscillating electric fields.
The principle of their operation relies only on the interaction between electric fields and point-like charges, and hence finds wide applicability, including to the confinement of single atoms \cite{Leibfried2003}, electrons \cite{matthiesenTrappingElectronsRoomTemperature2021}, molecules \cite{lohPrecisionSpectroscopyPolarized2013}, nanoparticles \cite{aldaTrappingManipulationIndividual2016} and microdiamonds~\cite{delordSpincoolingMotionTrapped2020}.
Systems consisting of a small number of trapped atomic ions have been employed as very accurate clocks \cite{brewerQuantumLogicClockSystematic2019}, for precision measurements, as sensors for small forces \cite{blumsSingleatom3DSubattonewton2018}, and are a leading platform for quantum information processing \cite{haffnerQuantumComputingTrapped2008} and networking \cite{moehringEntanglementSingleatomQuantum2007}.

Typically, a quadrupole potential oscillating in the radio-frequency range (\RF{}) provides effective time-averaged confinement in one (radial) plane, with a static (\DC{}) potential confining the particle along the remaining axis \cite{Leibfried2003}.
An inherent feature of such systems is \enquote{micromotion}: the coherent modulation of particle trajectories on the \RF{} time scale, which vanishes at the central null of the quadrupole potential.
If the equilibrium position of the particle does not coincide with this null, the particle undergoes excess micromotion that cannot be reduced by cooling.

Excess micromotion is rarely a desirable feature.
For atomic clocks using trapped ions, the modulation in velocity leads to an appreciable time-dilation shift in the observed frequency.
The associated uncertainty has dominated the accuracy budget of recent experiments \cite{brewerQuantumLogicClockSystematic2019}.
In quantum information experiments, micromotion can not only lead to increased heating of the secular modes of motion \cite{brouardHeatingTrappedIon2001}, but also introduces a periodic phase modulation to all interactions of the trapped ions with external fields (such as lasers or local microwave gradients).

Trap electrode geometries are usually designed such that the \RF{} and \DC{} fields both vanish at the intended trapping location.
For a single trapped particle, excess micromotion is thus the consequence of imperfections in the experimental realisation, such as \DC{} field offsets caused by stray charges on nearby elements of the trap apparatus.
To again shift the \DC{} field null onto the \RF{} null, static compensation potentials can be applied to a suitable combination of trap electrodes.

Various techniques to determine the required compensation fields have been proposed.
Some of these micromotion compensation methods make use of particularities of the experimental systems for which they were developed, for instance a co-trapped gas of ultra-cold atoms \cite{harterMinimizationIonMicromotion2013,meirDynamicsGroundStateCooled2016}, a high-finesse optical cavity collecting ion emission \cite{chuahDetectionIonMicromotion2013}, microwave near-field magnetic gradients \cite{warringTechniquesMicrowaveNearfield2013}, or fluorescence imaging optics with adjustable focus \cite{glogerIontrajectoryAnalysisMicromotion2015}.

Another, more generally applicable family of methods relies on micromotion-induced phase modulation of a laser beam in the ion rest frame \cite{berkelandMinimizationIonMicromotion1998}.
These include minimisation of the amplitude of the micromotion modulation sideband observed for a transition with a linewidth smaller than the trap drive frequency, or minimising temporal variations in the stimulated emission rate for a transition of linewidth comparable to or larger than the trap \RF{} (see \citeref{kellerPreciseDeterminationMicromotion2015} for a comprehensive discussion).

These methods measure the projection of micromotion onto the wavevector $\kvec{}$ of a probing laser.
As such, multiple probe beams are required to be able to resolve stray fields in all directions.
This requirement is a particular challenge for the planar \enquote{surface trap} designs produced through scalable microfabrication techniques.
There, it is usually not possible to achieve a direction $\kvec{}$ with an appreciable component perpendicular to the trap plane without the laser beam also striking the trap, which is generally undesirable due to charging effects (with the exception of the infrared beams necessary to cool some ion species \cite{allcockImplementationSymmetricSurfaceelectrode2010}).

\begin{figure}
	\includegraphics{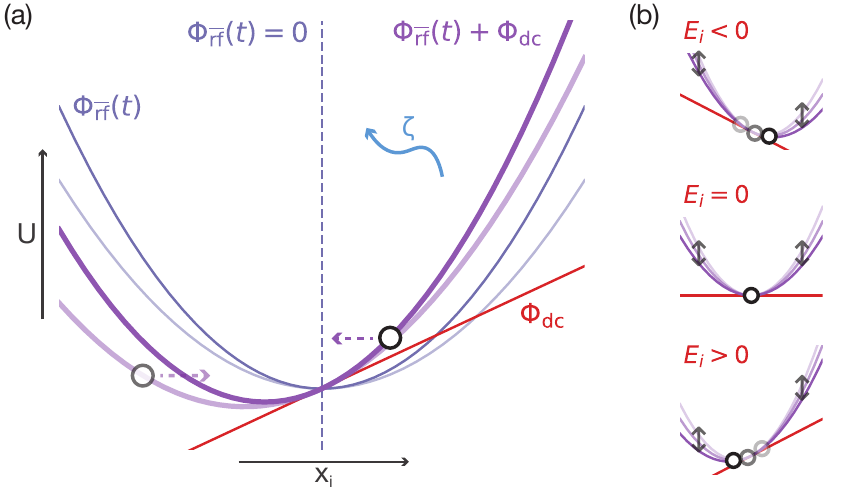}\\\hspace{12pt}
	\includegraphics{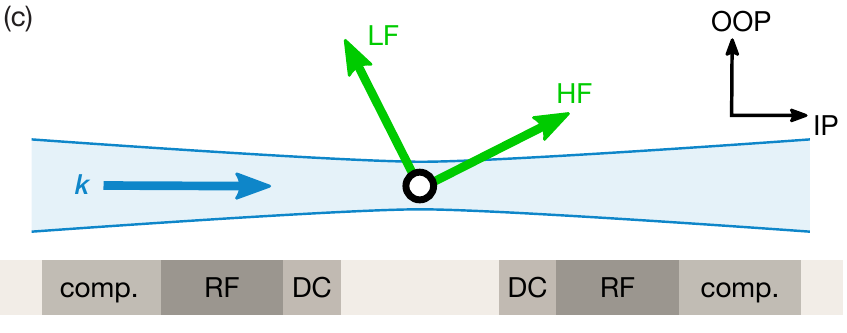}
	\caption{
		Principle: As illustrated in (a) for the direction $x_i$ along one mode of secular motion, amplitude modulation of the radio-frequency trap drive modulates the curvature of the \RF{}-induced pseudopotential $\Phi_{\overline{\mathrm{rf}}}(t)$.
		In the presence of a constant stray electric field $E_i$ (with potential $\dcpot$) this leads to coherent excitation of motion, whose amplitude reaches a steady state in the presence of some linear damping $\zeta$.
		This excitation vanishes and changes phase as the stray field passes zero, shown in (b).
		Diagram (c) shows a radial slice through a microfabricated surface ion trap, where all laser beams typically propagate in the plane parallel to the trap surface (along~$\kvec$).
		The two radial modes of motion, of respectively lower (LF) and higher (HF) frequency, span the plane perpendicular to the trap axis, and both have non-vanishing projection onto $\kvec$.
		By successively tuning the modulation frequency close to each of the modes, stray fields can thus be detected in both the in-plane (IP) and the out-of-plane (OOP) directions using only one laser beam.
	}
	\label{fig:principle}
\end{figure}

In this manuscript, we develop a method for complete stray field compensation requiring only a single laser beam, based on parametric excitation of the secular modes of motion of a trapped ion \cite{zhaoParametricExcitationsTrapped2002,ibarakiDetectionParametricResonance2011,narayananElectricFieldCompensation2011,tanakaMicromotionCompensationSurface2012,amiram.eltonySensitive3DMicromotion2013}.
As illustrated in \figref{fig:principle}, amplitude modulation of the \RF{} trapping potential can coherently excite secular motion if the \DC{} and \RF{} nulls do not coincide.
In the presence of dissipation, the steady-state motional amplitude is proportional to the component of the stray field in the respective mode direction.
The full shift of the equilibrium position from the \RF{} null can thus be determined by successively modulating the potential at frequencies close to that of every mode.
To quantify amplitude and phase of the resulting motion, we record and digitally demodulate the arrival times of photons scattered during laser cooling.
A single laser beam is sufficient for this as long as its wavevector $\kvec$ has a non-vanishing component along the direction of each mode.

As the applied compensation fields are varied, the amplitude of motion correlated to the excitation signal crosses zero at the point where the stray field vanishes, providing a robust signal for a micromotion compensation procedure.
The minimal degree of technical complexity associated with the proposed method makes it well-suited for integration into the micro-fabricated surface trap systems that represent the state of the art for quantum information processing and networking experiments using trapped ions.

In this manuscript, we first derive an analytic description of the method and its application to trapped ions in sections \ref{sec:theory} and \ref{sec:cooling}.
In \secref{sec:experimental-setup}, we then present experimental data from a \srplus{}/\caplus{} quantum network node \cite{stephensonHighrateHighfidelityEntanglement2020}, where we routinely compensate the stray fields in a surface trap to better than \SI{1}{\volt/\metre} in all directions.
Finally, we discuss some practical concerns regarding the use of this parametric excitation method to build a robust automated procedure for the compensation of stray fields in typical non-ideal surface traps – without detailed knowledge about the mode structure – in \secref{sec:compensation}.

\section{Parametric excitation}
\label{sec:theory}

\subsection{Equations of motion}

To describe the dynamics of a point-like particle of mass $m$ and charge $\charge$ in a Paul trap with drive frequency $\wRF$, it is convenient to expand both the static potential $\dcpot$ and the oscillating potential $\rfpot \cos(\wRF t)$ around the equilibrium position, which we assume, without loss of generality, to be in the coordinate origin. In the absence of stray fields, the respective gradients $\nabla \dcpot$ and $\nabla \rfpot$ vanish at this point, and the equations of motion for particle position $\vec{x} = (x_i)_{i = 1, 2, 3}$ are
\begin{equation}
	\ddot{\vec{x}} = -\frac{\charge}{2 m} \left(\hessian{\dcpot} + \hessian{\rfpot} \cos(\wRF t) \right) \vec{x} + O(|\vec{x}|^2),
\end{equation}
where $\hessian{(\cdot)}$ is the Hessian of the respective potential in the origin.
This can be recast as a coupled system of Mathieu-type equations,
\begin{equation}
	\ddot{\vec{x}} + \left(\frac{\wRF}{2}\right)^2 \left(\amat + 2 \qmat \cos(\wRF t)\right) \vec{x} = 0,
	\label{eq:matrix-mathieu}
\end{equation}
where the dimensionless matrices
\begin{equation}
	\amat = \frac{2 \charge}{m \wRF^2} \hessian{\dcpot}, \quad
	\qmat = \frac{\charge}{m \wRF^2} \hessian{\rfpot}
\end{equation}
are generalisations of the usual scalar Mathieu parameters $a$ and $q$.

If $\amat$ and $\qmat$ are simultaneously diagonalisable, the system is easily decoupled by a coordinate transformation $x_i \mapsto \tilde{x}_i$ into the shared eigenbasis, resulting in individual Mathieu equations
\begin{equation}
	\ddot{\tilde{x}}_i + \left(\frac{\wRF}{2}\right)^2 \left(\tilde{a}_i + 2 \tilde{q}_i \cos(\wRF t)\right) \tilde{x}_i = 0,\quad  i = 1, 2, 3,
	\label{eq:decoupled-mathieu}
\end{equation}
with stability parameters $\tilde{a}_i = 2 \charge / (m\, \wRF^2)\ \partial^2\dcpot/\partial\tilde{x}_i^2$ and $\tilde{q}_i = \charge / (m\, \wRF^2)\ \partial^2\rfpot/\partial\tilde{x}_i^2$.

This case is widely studied \cite{Leibfried2003}, particularly as both $\dcpot$ and $\rfpot$ are well-approximated by electrode-aligned quadrupole potentials in macroscopic linear (\enquote{four-rod}) Paul traps.
In symmetric surface trap designs, on the other hand, the static voltages are usually chosen precisely such that $\dcpot$ induces a rotation of the secular modes of motion away from the principal axes of $\rfpot$ to ensure a non-zero projection onto the cooling laser beams.
We still restrict the following discussion to the scalar case for simplicity, as to obtain an analytical approximation, we apply the pseudo-potential approximation anyway, where both \DC{} and the time-averaged \RF{} curvature are combined into a single confining potential.
If necessary for quantitative calculations, Floquet solutions to the general \eqnref{eq:matrix-mathieu} could similarly be found using a technique of infinite-continued matrix inversions \cite{houseAnalyticModelElectrostatic2008,landaClassicalQuantumModes2012}.

In particular, we will consider the case of a laser-cooled particle in a trap with an additional, unwanted electric field of uniform strength $\vec{E} = (E_i)_i$, which we wish to probe by amplitude-modulating the driving RF field at frequency $\wAM$ and modulation depth $\h$.
Dropping tildes and mode index subscripts, \eqnref{eq:decoupled-mathieu} then becomes
\begin{equation}
\begin{gathered}
	\left(\frac{\wRF}{2}\right)^2 \Big(a\, + 2 q \big(1 + \h \, \cos(\wAM t) \big) \cos(\wRF t)\Big) \, x \, +\\
	2 \zeta\,\dot{x} + \ddot{x} = \frac{E\, \charge}{m},
\end{gathered}
\label{eq:decoupled-mathieu-damped}
\end{equation}
where $\zeta$ gives the strength of linear viscous damping induced by laser cooling.

\subsection{Pseudopotential approximation}

To understand the effect of the amplitude-modulated trapping potential on the ion's secular motion, it is instructive to consider \eqnref{eq:decoupled-mathieu-damped} in the pseudopotential approximation, where fast dynamics on the scale of $\wRF$ are ignored.
Here, for $|a| \ll |q| \ll 1$ and $\wAM, \zeta \ll \wRF$, we obtain
\begin{equation}
	\ddot{\bar{x}} + 2 \zeta \dot{\bar{x}} + \left(\frac{\wRF}{2}\right)^2 \left(a + \frac{q^2}{2} \big(1 + \h \cos(\wAM t)\big)^2\right) \bar{x} = \frac{E\, \charge}{m}
\end{equation}
for the slowly-varying position $\bar{x}$. Defining
\begin{equation}
	\wm \defeq \frac{\wRF}{2}\ \sqrt{a + \frac{q^2}{2} \left(1 + h^2 / 2\right)} \label{eq:pseudopot-modefreq}
\end{equation}
for the resulting frequency of secular motion and
\begin{equation}
	\hred \defeq \frac{h}{1 + \h^2 / 2} \left(1 - \frac{a \wRF^2}{4 \wm^2}\right)
\end{equation}
as an \enquote{effective} modulation index, reflecting the fact that the contribution of $\dcpot$ to the radial curvature is static, we obtain
\begin{equation}
	\ddot{\bar{x}} + 2 \zeta \dot{\bar{x}} + \wm^2 \left(1 + 2 \hred \cos(\wAM t) + \h \hred \cos(2 \wAM t) \right) \bar{x} = \frac{E\, \charge}{m}.
	\label{eq:pseudopotential-damped}
\end{equation}
Neglecting the $O(\h^2)$ term oscillating at twice the excitation frequency (solutions up to second order in $h$ are given in appendix~\ref{sec:analytical-mathieu-solns}), this can be written as
\begin{equation}
	\ddot{\bar{x}} + 2 \zeta \dot{\bar{x}} + \left(\frac{\wAM}{2}\right)^2 \left(A + 2 Q \cos(\wAM t)\right) \bar{x} = \frac{E\, \charge}{m},
	\label{eq:pseudopotential-damped-mathieu}
\end{equation}
another damped Mathieu equation with stability parameters $A \defeq (2 \wm / \wAM)^2$ and $Q \defeq \hred A$. Note that the parameter regime is quite different from the case describing ion motion in the radial plane, i.e. \eqnref{eq:decoupled-mathieu}, where $|a| \ll |q|$ is usually a small amount of anticonfinement resulting from the axially confining \DC{} potential, and $|q| \ll 1$.
Here, however, we are interested in the regime close to the onset of parametric instability, that is, $A \approx \prindex^2 > 0$ for the resonance index $\prindex \in \mathbb{N}$, or $\wAM \approx 2\, \wm / \prindex$.

Without damping and if $\wAM = 2\, \wm / \prindex$ exactly, the system is unstable for any non-zero excitation amplitude $h$, and solutions are oscillatory with an exponentially growing amplitude over time.
Parametric instabilities of this kind have been studied in a wide variety of systems \cite{rajasekarParametricResonance2016}, notably for mass-selective removal of ions from Paul traps \cite{schmidtMassselectiveRemovalIons2020}.
Previous work on stray field compensation using parametric excitation \cite{ibarakiDetectionParametricResonance2011,narayananElectricFieldCompensation2011,tanakaMicromotionCompensationSurface2012} has also made use of the fact that these quickly-growing solutions can easily lead to large ion orbits where changes to the average fluorescence scattering rate of the ion during laser cooling can be observed.
For that purpose, excitation at the first parametric resonance ($\prindex = 1$, i.e.~$\wAM = 2 \, \wm$) is most convenient, as the stability can be shown to be lowest there.

In contrast, here (as in ref.~\cite{amiram.eltonySensitive3DMicromotion2013}) we are interested in the second parametric resonance where $\wAM \approx \wm$, but for a small modulation depth $h$ such that the damping term suppresses the exponential growth in amplitude.
As shown in the next section, solutions then tend towards oscillation at $\wAM$ if a stray field $E \neq 0$ is present, with a well-defined phase relationship to the amplitude-modulation signal independent of the initial conditions.

\subsection{Steady-state solutions}
\label{sec:steady-state-soln}

We now derive analytical approximations for the ion motion resulting from parametric excitation, as modelled by \eqnref{eq:pseudopotential-damped}.
Solutions differ in character depending on the relative strength of excitation and damping.
First, consider the homogeneous case with no stray field, ${E = 0}$.
The stability of solutions can then be investigated using standard techniques for the treatment of Mathieu equations (see appendix~\ref{sec:analytical-mathieu-solns}).
For small amounts of damping, $\zeta \ll \wm$, the second parametric resonance is excited most strongly if $\wAM = \wm$, where the stability condition can be approximated as
\begin{equation}
	\hred \ltish \frac{2 \zeta}{\wm}
\end{equation}
to first order in the effective modulation index $\hred$.

In this region of parameter space, solutions for arbitrary initial conditions decay to the equilibrium at $\bar{x}= 0$ at a characteristic rate given by the damping strength $\zeta$. It can be shown \cite{slaneAnalysisPeriodicNonautonomous2011} that the presence of an inhomogeneous term $E \neq 0$ does not alter the stability considerations. However, rather than tending towards a stationary equilibrium, solutions will then generally exhibit bounded steady-state behaviour.

Such steady-state solutions can be found using Floquet theory, which suggests an ansatz of the form
\begin{equation}
	\bar{x}(t) = \sum_{n=-\infty}^{\infty} c_n\, \ee^{\ii n \wAM t}
	\label{eq:ansatz}
\end{equation}
with fixed coefficients $(c_n)_{n \in \mathbb{Z}} \subset \mathbb{C}$, $c_n = \left(c_{-n}\right)^{*}$. Inserting this into \eqnref{eq:pseudopotential-damped-mathieu} yields a set of recurrence relations for the coefficients as detailed in appendix~\ref{sec:analytical-mathieu-solns}. To first order in $\hred$ ($\hred^2 \ll 1$), all terms with $|n| > 1$ vanish, and in
\begin{equation}
	c_0 = \frac{E \charge}{m\, \wm^2}
\end{equation}
we recover the expected shift of the average ion position due to the external field.
Introducing the normalised excitation frequency and damping parameters
\begin{equation}
	f \defeq \frac{\wAM}{\wm}, \quad d \defeq \frac{\zeta}{\wm},
\end{equation}
and defining $C \defeq 2 c_{-1}$ for compactness of notation such that $\bar{x}(t) \approx c_0 + \operatorname{re}(C\, \ee^{-\ii \wAM t})$, we obtain
\begin{equation}
	|C|^2 = \frac{(2 c_0 \hred)^2}{(f^2 - 1)^2 + 4 f^2 d^2}, \quad \tan\arg(C) = \frac{2 f d}{f^2 - 1}
	\label{eq:correlated-ampl-phase}
\end{equation}
to second order in $\hred$.

In the limit of small modulation depth, the system thus behaves like a damped, driven harmonic oscillator.
If the mode is excited exactly on resonance, the resulting motion is (for $c_0 < 0$) phase-shifted by $\pi / 2$ compared to the applied amplitude modulation
At modulation frequencies much smaller or much larger than the mode frequency the shift is $0$ or $\pi$, respectively, with the width of the transition region being proportional to the damping coefficient.

\begin{figure}
	\includegraphics[trim={0 0.6cm 0 0}]{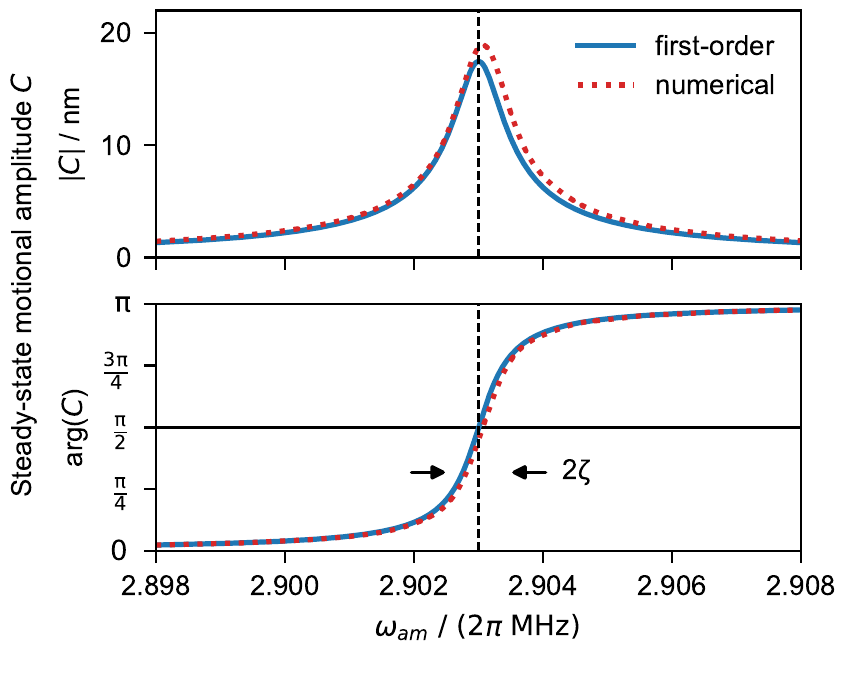}
	\caption{
		Predicted amplitude and phase of steady-state oscillation following excitation at $\wAM$ with modulation index $\h = 0.007$ and damping $\zeta = \SI{2.4}{\milli\second^{-1}}$ in the presence of a stray field with $E \charge / m = \SI{-1}{m / s^2}$, for a motional mode described by $a = -0.00259$, $q = 0.179$, and $\wRF = 2 \pi \cdot \SI{50}{\MHz}$ ($\wm = 2 \pi \cdot \SI{2.903}{\MHz}$).
		The solid blue curves show the result obtained to first order in $\h$ in the pseudopotential approximation, given by \eqnref{eq:correlated-ampl-phase}.
		The dashed red curves were obtained by numerically integrating the full equations of motion.
		For ease of comparison, the latter were shifted horizontally by the difference between the mode frequency obtained from full simulations without modulation and the Dehmelt approximation $\wm \approx \wRF / 2 \, \sqrt{a + q^2 / 2}$, which is indicated by the vertical dashed lines.
	}
	\label{fig:theory-freq-scan}
\end{figure}

Figure~\ref{fig:theory-freq-scan} shows the behaviour predicted by \eqnref{eq:correlated-ampl-phase} for a set of parameters typical for a microfabricated surface trap (see \secref{sec:experimental-setup}), along with numerical simulations for the full ion dynamics as per \eqnref{eq:decoupled-mathieu-damped}.
For the small modulation depth of $\h = 0.007$ used here, the approximation from \eqnref{eq:correlated-ampl-phase}, which includes only terms linear in $h$, only slightly underestimates the amplitude near resonance.

Crucially, the correlated amplitude $C$ is linear in $c_0$, and thus in the stray field strength $E$.
This linear dependence enables a simple stray field compensation procedure:
for an arbitrary fixed modulation frequency $\wAM$ close to the frequency of one mode of motion, measure $C$ for a variety of static compensation fields; its zero-crossing then marks the compensated point where the stray field is exactly cancelled.
In principle, excitation exactly on resonance would give the largest signal, but at this point, the observed amplitude would be very sensitive to small deviations in the mode frequency, as caused for instance by trap anharmonicity or technical drifts in the \RF{} voltage level.
In practice, detuning $\wAM$ from $\wm$ by a small amount on the order of $\zeta$ thus increases the robustness of the procedure.
While larger detunings can also be used, this also increases the modulation depth required to reach a given sensitivity, eventually leading the approximations made here ($\h \ll 1$) to break down.

\section{Approximate ion dynamics in the linearised two-level picture}
\label{sec:cooling}

To connect the results obtained in the last section to the experiment, it is necessary to model the internal degrees of freedom of the trapped ion, both in order to extract the damping coefficient $\zeta$ resulting from laser cooling, and to derive the modulation in fluorescence count rate used to infer the motional amplitude.
The simplest adequate model is that of a two-level system always assumed to be in equilibrium (that is, with internal state dynamics fast compared to $\wm$).


Consider a two-level system with ground state $g$ and excited state $e$, connected by a transition with linewidth $\Gamma$, and driven to equilibrium by an external field (here, the cooling laser) with Rabi frequency $\Omega_{eg}$ and a detuning $\Delta$ from resonance.
The rate of photons scattered is
\begin{equation}
	R = \Gamma \, \rho_{ee}(\Delta),
\end{equation}
where the excited state population $\rho_{ee}$ can be written as
\begin{equation}
	\rho_{ee}(\delta) = \frac{s / 2}{1 + s + \left(\frac{2\delta}{\Gamma}\right)^2}
\end{equation}
in terms of the saturation parameter $s\, \defeq 2\, \Omega_{eg}^2 / \Gamma^2$ \cite{Leibfried2003}.

Due to the Doppler shift, ion motion gives rise to an instantaneous detuning $-k \dot{x}$, where $k$ is the projection of the wavevector of the laser beam onto the mode direction.
Assuming the ion velocity changes slowly, such that the internal state is in equilibrium at any point, we can write
\begin{equation}
	R = \Gamma \left(\rho_{ee}(\Delta) - \frac{\partial \rho_{ee}}{\partial \delta}\Bigr|_{\delta=\Delta} k \dot{x} + O\big((k \dot{x} / \Gamma)^2 \big)\right).
\end{equation}
Neglecting momentum diffusion from the isotropically emitted photons as well as any temporal correlations between absorption and emission events, laser cooling can then be modelled through the average scattering force $F = \hbar k R$, giving the linear damping term as $\zeta = -1/(2 m)\ \partial F / \partial \dot{x}$.

\label{sec:detecting-motion}

If a fraction of the emitted photons are registered by some detection system with overall efficiency $\eta$, motion excited by the trap potential modulation, such that $\dot{\bar{x}}(t) = \wAM \im{C \ee^{-\ii \wAM t}}$ for the complex amplitude $C$ defined in \secref{sec:steady-state-soln}, leads to a corresponding modulation in the observed count rate $r(t) = \eta R(t)$:
\begin{equation}
	r(t) \approx  \eta \left(R_0 + \frac{4 k \wAM}{\Gamma} \frac{s \Delta}{\left(1 + s + \left(\frac{2\Delta}{\Gamma}\right)^2\right)^2} \im{C \ee^{-\ii \wAM t}}\right),
	\label{eq:correlated-rate-model}
\end{equation}
where $R_0 = R(\dot{x} = 0)$ is the average scattering rate, which coincides with the unexcited case.

As such, by measuring the frequency component of the count rate at the excitation frequency, which we call the correlated rate $S$, we can recover the excitation amplitude via
\begin{equation}
	S \defeq \left\langle r(t) \ee^{\ii \wAM t} \right\rangle_t =
	-\ii \frac{2 \eta k \wAM}{\Gamma} \frac{s \Delta}{\left(1 + s + \left(\frac{2\Delta}{\Gamma}\right)^2\right)^2}\, C,
	\label{eq:correlated-rate-demod}
\end{equation}
where $\left\langle \cdot \right\rangle_t$ denotes the average over an integer multiple of excitation periods.

In the experiment, $S$ is easily estimated by recording the arrival times $(t_n)_{n = 1, \ldots, N}$ of $N$ photons over some time $T \gg 2 \pi / \wAM$ and digitally demodulating the signal as
\begin{equation}
	\left\langle r(t) \ee^{\ii \wAM t} \right\rangle_t \approx \frac{1}{T} \sum_{n=1}^{N} \ee^{\ii \wAM t_n}.
	\label{eq:timestamp-demodulation}
\end{equation}
The finite number $N$ of observed photons leads to a statistical variance, i.e.~photon shot noise, of $N / (2 T^2) = \eta R_0 / (2 T)$ per complex component of $S$.
$S$ is linear in the motional amplitude $C$ and thus, as shown in the previous section, in the stray electric field $E$.
If we use this procedure to estimate $E$, the statistical variance for an observation time $T$ is
\begin{equation}
	{\sigma_E}^2 = \underbrace{\left|\frac{E}{S}\right|^2 \frac{\eta R_0}{2}}_{=: \alpha^2} \frac{1}{T},
\end{equation}
where we have introduced the overall (inverse) sensitivity $\alpha$ of the stray field measurement (unit: $\si{\volt \meter^{-1}} / \sqrt{\si{\hertz}}$).
Substituting the concrete expressions for $S$ and $C$ obtained earlier, we thus obtain the sensitivity of the method in this linearised model as
\begin{equation}
	\begin{gathered}
	\alpha^2 \defeq \left(\frac{m}{2 \hred \charge}\right)^2 \left((f^2 - 1)^2 + 4 f^2 d^2\right) \cdot \\ \frac{\Gamma^3 \left(1 + s + \left(\frac{2 \Delta}{\Gamma}\right)^2\right)^3}{16 \eta k^2 \Delta^2 s},
	\end{gathered}
	\label{eq:field-error}
\end{equation}
noting that the laser parameters $s$ and $\Delta$ also enter the normalised damping parameter $d$.

Without further constraints, attempts to choose experimental parameters for optimal sensitivity by minimising \eqnref{eq:field-error} are ill-fated.
For excitation exactly at the mode frequency ($f = 1$), reducing the damping coefficient would lead to arbitrarily large ion orbits and thus relative signal rates $S / (\eta R_0)$.
In reality, however, nonlinearities of trap and cooling forces \cite{akermanSingleionNonlinearMechanical2010} and the onset of parametric instability limit the achievable amplitudes.
The simple, linear model derived here is thus inadequate to derive an ultimate bound on the sensitivity of this method.
Nevertheless, if the allowable motional amplitude $|C|$ is fixed (for instance, as limited by trap uniformity), \eqnref{eq:field-error} attains a minimum at $-\Delta = \Omega_{eg} = \Gamma / 2$.
Similarly, if $\hred$ and $f$ are fixed (choosing, for instance, a detuning $1 - f \approx 5 \cdot 10^{-3}$ to decrease sensitivity to trap frequency drifts), the sensitivity is again found to be maximal near $\Delta \approx - \Gamma / 2$ over a range of mode frequencies and linewidths typical of trapped ion experiments.
This suggests $\Delta = -\Gamma / 2$ to be a suitable starting point for further empirical optimisation.

\section{Experimental setup}
\label{sec:experimental-setup}

We demonstrate this method using a single \srplus{} ion in an \enquote{HOA2} microfabricated surface trap manufactured by Sandia National Laboratories \cite{maunzHighOpticalAccess2016}.
A single \srplus{} ion is trapped at an \RF{} null designed to be \SI{68}{\micro\metre} above the surface.
Atoms are loaded from a thermal source~\cite{ballanceShortResponseTime2018} through a two-stage resonant photoionisation process using laser beams at \SI{461}{\nano\metre} and \SI{378}{\nano\metre}, which propagate parallel to the trap plane and are focussed at the trap centre, as are the laser cooling beams.
The $5s\, {S}_{1/2} \leftrightarrow 5p\, {P}_{1/2} \leftrightarrow 4d\, {D}_{3/2}$ cooling cycle is addressed by \SI{422}{\nm} and \SI{1092}{\nm} beams (of approximately $\pi$ and $\sigma^\pm$ polarisations) perpendicular to a \SI{0.5}{\milli\tesla} magnetic field, each oriented at \SI{45}{\degree} to the trap axis.
The power of each beam is stabilised by sampling a small portion of the beam onto a photodiode.
Ion fluorescence is collected through a \SI{60}{\micro\metre} slot in the trap centre and imaged onto a photomultiplier tube, leading to an overall collection efficiency of $\eta \approx 0.4\%$ with negligible background count rate.
A real-time control system built on the ARTIQ/Sinara open-source ecosystem \cite{ARTIQ} records photon arrival times with \SI{1}{\nano\second} precision.

The electrode arrangement is similar to \figref{fig:principle}.
The \RF{} electrodes are resonantly matched to \SI{50}{\ohm} at \SI{50.2}{\MHz} using a single-stage lumped-element $LC$ circuit with a quality factor of approximately $Q \approx 40$, for a resulting \RF{} amplitude of $\approx \SI{162}{\volt}$.
Amplitude modulation is applied using a direct digital synthesis source (DDS) with phase referenced to the same clock used for input timestamping, as described in appendix~\ref{sec:rf-setup}.
The \DC{} electrodes are driven by a multi-channel 16-bit, $\SI{\pm 10}{\volt}$ digital-to-analogue converter (DAC), with voltages chosen to result in typical mode frequencies of $2 \pi \times 1.85 \si{\MHz}$, \SI{2.28}{\MHz} and \SI{2.90}{\MHz}, respectively, for the axial and two radial modes.
A static quadrupole term is added such that the radial mode vectors are rotated by \SI{28}{\degree} about the trap axis, resulting in overall angles of \SI{70}{\degree} and \SI{51}{\degree} between the cooling beams and the low- and high-frequency radial modes, respectively.

The sensitivity of the electric field at the ion position to the voltages applied to the global compensation electrodes in the in-plane and out-of-plane directions is \SI{0.5}{\volt \metre^{-1} / \milli\volt} and \SI{0.6}{\volt \metre^{-1} / \milli\volt}, respectively, such that the DAC resolution limits the stray field compensation precision to \SI{0.2}{\volt \metre^{-1}}.

\begin{figure}
	\includegraphics{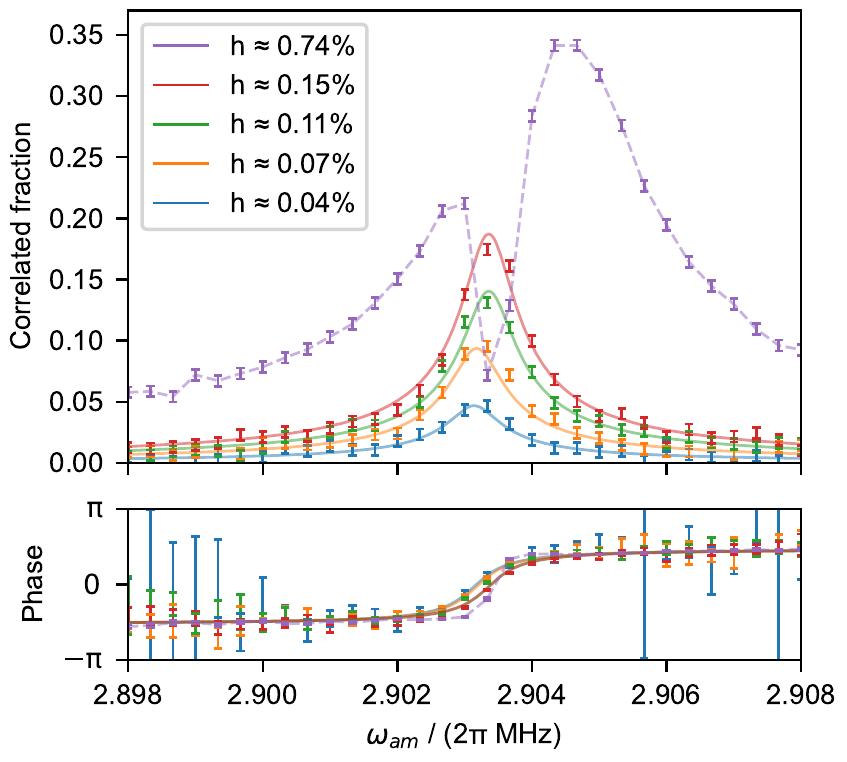}
	\caption{
		Observed amplitude and phase of the correlated photon count rate versus the modulation frequency.
		The signal amplitude $S$ is given as a fraction of the approximately constant total count rate of $\eta R_0 = \SI{15}{\kilo\hertz}$, for cooling parameters $\Delta = -2\pi \times \SI{13}{\mega\hertz}$, $s \approx 0.9$.
		The excitation frequencies are near the higher-frequency radial mode, and we intentionally apply an offset field of $\SI{5}{\volt/\metre}$ along the mode direction.
		Data for five different modulation indices $h$ are shown:
		For the smaller four, the solid lines lines show a fit of the model from \eqnref{eq:correlated-ampl-phase}, allowing for an individual offset in mode frequency and a constant phase shift common to all curves (yielding a damping rate $\zeta = \SI{2.4}{\milli\second^{-1}}$).
		For excitation at $h \approx \SI{0.74}{\percent}$, where nonlinear effects dominate, the dashed line connects the points to guide the eye.
		Error bars show $1 \sigma$ statistical errors from photon shot noise (\num{50000} photons per point).
	}
	\label{fig:freq-scan}
\end{figure}

Figure~\ref{fig:freq-scan} shows amplitude and phase of the correlated scattering rate signal $S$ as a fraction of the total fluorescence rate for excitation near the higher-frequency radial mode, with the ion shifted from the \RF{} null by $\SI{5}{\volt / \metre}$ along the direction of that mode.
For these data, the ion is continuously cooled.
After switching on the amplitude-modulation signal, a \SI{300}{\micro\second} settling period reduces the influence of the initial conditions.
Timestamps are then recorded for each PMT click until a pre-defined number of photons (here \num{50000}) has been observed.
The signal is demodulated according to \eqnref{eq:timestamp-demodulation} by comparing the timestamps to the phase origin of the parametric excitation envelope signal \footnote{There is a non-zero total delay through the \RF{} modulation and detection chain, so a linear phase equivalent to equivalent to \SI{1.8}{\micro\second} was subtracted from the data for display purposes, recovering an asymptotically flat phase response far off resonance. This does not affect the stray field compensation algorithm described in \secref{sec:compensation}.}.

\begin{figure*}
	\includegraphics[trim={0 0.3cm 0 0}]{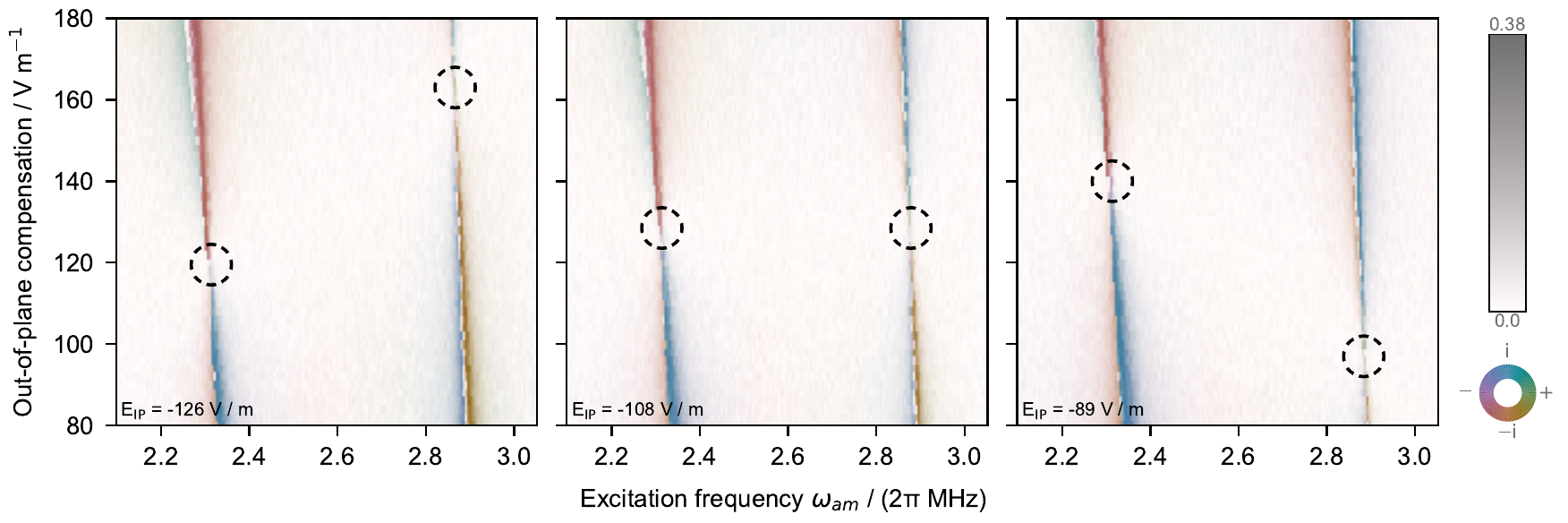}
	\caption{Complex fractional amplitude of correlated photon arrival times, as a function of excitation frequency and applied field normal to the trap surface, for radial in-plane compensation fields of $E_{\mathrm{IP}} = \SI{-126}{\volt / \metre}$, \SI{-108}{\volt / \metre}, and \SI{-89}{\volt / \metre}, respectively. Colour shows phase, and brightness shows amplitude (extracted from \num{10000} photons observed per point). The dashed circles highlight the points where the projection of micromotion onto the respective radial mode vanishes. Only when $E_{\mathrm{IP}}$ is chosen such that the stray field in the in-plane direction is minimised (centre plot) can both modes be compensated. The dependence of radial mode frequencies on ion position due to trap anharmonicity is visible as a slant to the resonance lines.}
	\label{fig:freq-oop-scans}
\end{figure*}

The behaviour for a number of different excitation amplitudes is shown, with cooling parameters $\Delta = -2\pi \times \SI{13}{\mega\hertz}$, $s \approx 0.9$.
The absolute scaling of the depth of the applied modulation after the trap resonator was calibrated to about 1 s.~f.~by probing the signal at a capacitive voltage divider after the resonant circuit with an oscilloscope.
For small modulation indices, the behaviour agrees with the linear model derived in \secref{sec:steady-state-soln}, shown in the form of fit curves, with mode frequency, damping rate, and overall amplitude factor as free parameters.
The observed rate grows linearly with modulation amplitude (and distance from the \RF{} null, which is constant here) until non-linearities in trap and motional coupling start to become significant.
Empirically, we observe a gradual saturation behaviour towards a correlated fraction of $0.35$ of the total fluorescence rate, i.e.~a scattering rate amplitude modulation depth of $0.7$.
The observed mode frequencies systematically increase towards higher excitation amplitudes.
The origin of this effect is unclear; while \eqnref{eq:pseudopot-modefreq}, derived in the pseudopotential approximation, does predict a (larger) shift, this is not borne out by numerical simulations of \eqnref{eq:decoupled-mathieu-damped} (see appendix~\ref{sec:mode-freq-shift}).

For larger excitation amplitudes, a sharp drop in correlated photon rate is observed near resonance.
This effect cannot be reproduced by the linear response model; we attribute it to ion response nonlinearities (when $|k \dot{\bar{x}}|$ reaches a significant fraction of the linewidth), trap nonlinearities (for large orbits), and the onset of parametric instability.
Empirically, we observe that for excitation near $\wm$ the ion remains confined in the trap even under those conditions.
Higher excitation amplitudes at larger detunings can be a useful working point for the stray field compensation algorithm discussed in the following section, as the effectively broadened response reduces the sensitivity to changes in mode frequency.

\section{Stray field compensation}
\label{sec:compensation}

By probing for parametric excitation near each of the mode frequencies, we can measure the stray field component along each mode.
Thus, stray fields in all spatial directions can be compensated.
As in most linear Paul traps, the \RF{} potential curvature along the trap axis is very small in our system by design, so we present data only for the plane spanned by the two radial modes in the following.

As illustrated in \figref{fig:freq-oop-scans}, this is a multi-dimensional optimisation problem.
Even when only varying the applied compensation fields along a single direction, there will usually be a point for each mode where the excitation vanishes, i.e.~the shift between \DC{} and \RF{} minimum along that mode direction is zero.
However, unless the spatial orientation of the motional modes is known to high precision so that an appropriate compensation voltage basis can be chosen (taking into account, for instance, the contributions of any stray quadrupole potentials), such a one-dimensional search will in general not yield enough information to determine the voltage set that positions the ion exactly on the \RF{} null.

\begin{figure*}
	\includegraphics{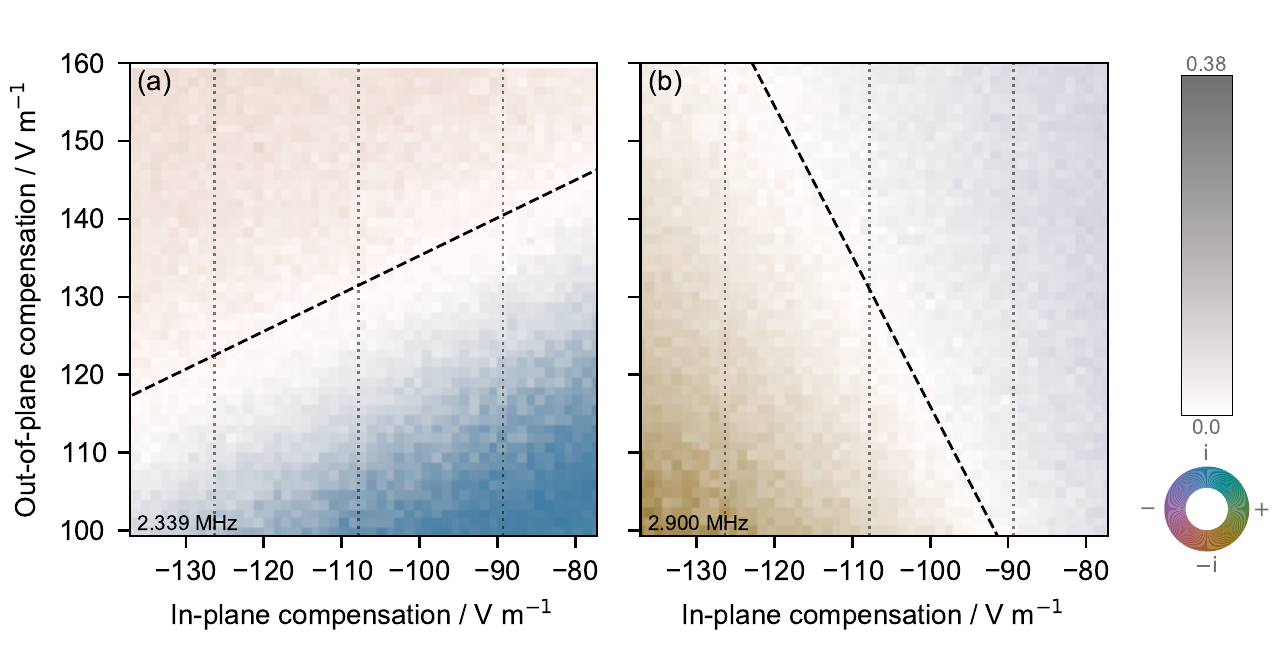}\hfill\includegraphics[trim={17mm 0 0 0}]{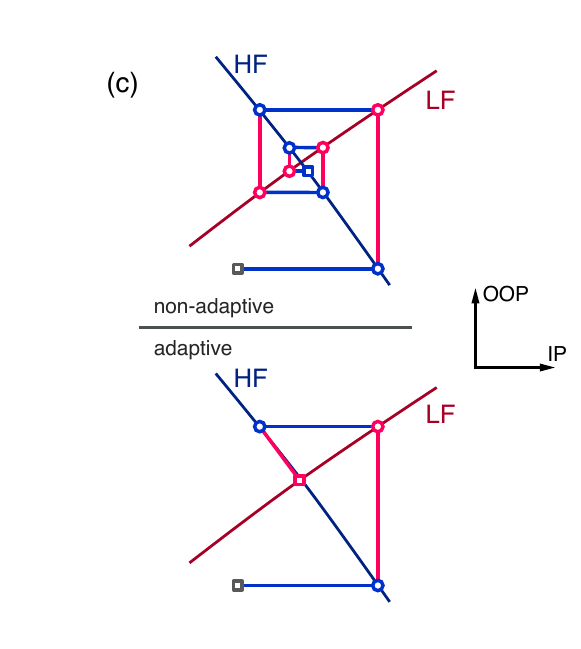}
	\caption{
		Stray field compensation in two dimensions.
		\textbf{(a), (b): }Complex fractional amplitude of photon arrival times correlated with the fixed excitation frequency of \SI{2.339}{\mega\hertz} and \SI{2.900}{\mega\hertz}, respectively, for a detuning of \SI{22}{\kilo\hertz} relative to each radial mode in the scan centre.
		Along the dashed lines, the residual displacement of in the direction of the given mode is minimised; all radial micromotion is compensated in the centre of the plot, where both lines would intersect.
		The dotted vertical lines mark the values for $E_{\mathrm{IP}}$ used in the three scans shown in \figref{fig:freq-oop-scans} for visual comparison.
		Both mode frequencies increase towards negative out-of-plane fields, reducing the detuning, giving rise to the stronger signals in the lower part of the plots.
		\textbf{(c): }To simultaneously compensate stray fields in both directions, a fixed-point iteration scheme can be employed, where the excitation on either mode is minimised in an alternating fashion by applying compensation fields in some pre-defined basis (here in the IP and OOP directions).
		This simple strategy is successful as long as the mode orientation is known to better than \SI{45}{\degree} (top), but an adaptive algorithm, which at each point estimates the local mode directions from previous measurements to avoid moving the ion off the respective other null, can greatly speed up convergence (bottom).
	}
	\label{fig:mode-directions}
\end{figure*}

This is further illustrated in \figref{fig:mode-directions} (a) and (b), which shows the motional amplitudes observed as the compensation fields vary in both radial directions for a fixed excitation frequency.
For each mode, there exists a linear minimum in excitation amplitude, where the projection of stray field onto the mode direction vanishes, and the ion is merely translated perpendicular to the mode direction.
A linear fit to each dataset reveals mode directions of \SI{-63}{\degree} and \SI{26}{\degree} relative to the trap plane (based on the geometry of the compensation electrodes), in reasonable agreement with the \SI{28}{\degree} intended mode axis rotation angle.
At the point where both lines intersect, the trap is compensated: the ion is centered on the \RF{} null and no parametric excitation is observed.

Acquisition of two-dimensional data of this nature is unnecessarily time-consuming.
As illustrated in \figref{fig:mode-directions} (c), as long as the direction of the applied compensation fields relative to the normal modes is known to better than \SI{45}{\degree}, a simple fixed-point iteration method, where excitation for each mode is minimised in an alternating fashion, will converge to the compensation optimum.
The orientation of each mode, expressed in terms of compensation voltages, is typically constant to good approximation, at least close to the compensated point.
If, at each step, the direction of the currently inactive mode(s) is estimated from previous iterations, and the probe direction is then chosen not to displace the ion in that direction, convergence can be greatly accelerated.
Note that the obtained motional mode directions are not necessarily normal when expressed in the compensation voltage basis, for instance if the fields generated by each electrode voltage set are not in fact orthogonal at the ion position, or of different magnitude.

In practice, a further complication arises from the fact that the observed mode frequencies will shift slightly as a function of the applied compensation voltages.
One reason for this is that the \DC{} and \RF{} potential curvatures are not necessarily constant over the volume probed by the ion positions, especially in surface electrode traps.
The potentials introduced by the compensation electrodes are also not in general curvature-free.
In \figref{fig:freq-oop-scans}, this shift appears as a slant to the resonance features.
To some extent, this can be mitigated by choosing a parametric excitation frequency far enough detuned from resonance that the change in response remains acceptably small over the region of interest.
However, if the region of compensation fields to be investigated is large, the detuning required for this might be too big to be practical, given that the achievable damping rate and amplitude modulation index will be limited.

\begin{figure}
	\centering
	\includegraphics{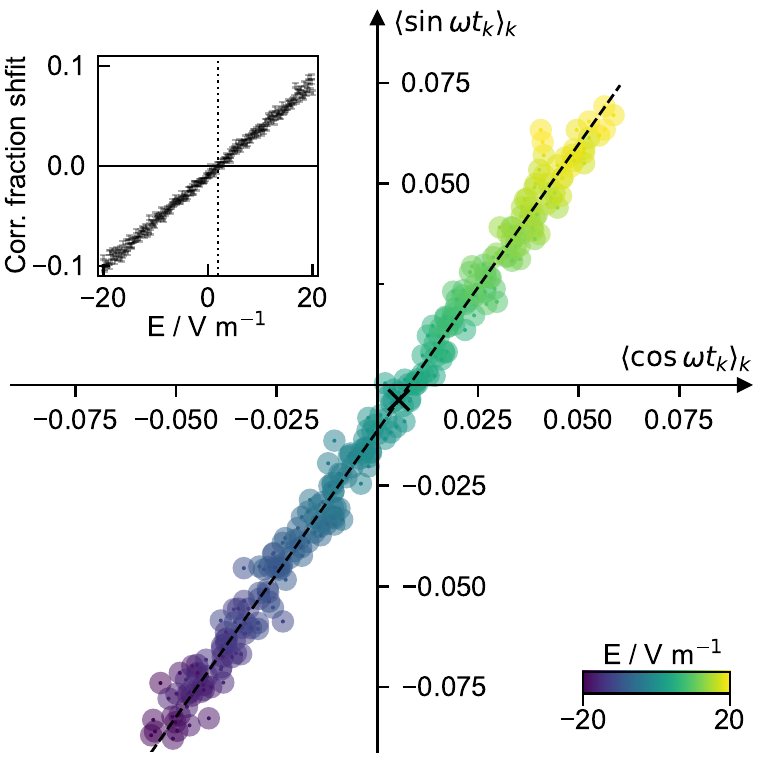}
	\caption{
		Correlated scattering rate signal for excitation close to a \SI{2.88}{\mega\hertz} radial mode (detuning $-\SI{22}{\kilo\hertz}$, $\h \approx 0.02$),  given as fraction of the \SI{30}{\kilo\hertz} total count rate for varying compensation fields along the mode direction.
		The dots are coloured to denote the applied fields, ranging from \SI{-20}{\volt / \metre} (dark) to \SI{20}{\volt / \metre} (light) relative to an arbitrary value close to the compensation optimum; their size corresponds to the $1 \sigma$ statistical error from \num{100000} observed photons each.
		The dashed line shows a linear complex least-squares fit to the data as a function of $E$-field.
		In the inset the same data is shown in one-dimensional form, with the vertical axis now giving the projection of each point onto the best-fit line, where zero corresponds to the point on the line closest to the coordinate origin.
		The symbols show the statistical error, confirming that the magnitude of excitation is linear to excellent approximation.
		The cross in the main plot denotes the inferred compensation optimum, that is, the point on the fit line closest to the origin.
		The same point, corresponding to $E = \SI{1.95(3)}{\volt / \metre}$, is indicated in the inset by the dotted line.
	}
	\label{fig:closeup}
\end{figure}

To robustly find a set of compensation electrode voltages without prior knowledge of the mode structure, we thus employ a multi-step algorithm (still alternating between modes).
Given a frequency estimate for the modes of interest and an initial guess and search radius for the respective compensation voltage sets to apply, iterate over the following steps:
\begin{enumerate}
	\item Determine the frequency of the mode in question at the current compensation settings plus a shift of one search radius in one direction. This can for instance be achieved by varying the excitation frequency and fitting a step of $\pi$ in phase to the observed correlated amplitude (see~\figref{fig:freq-scan}).
	\item Repeat step 1 for voltages shifted by one radius in the opposite direction, thus obtaining a linear model for mode frequency as a function of compensation voltage.
	\item Then, vary the applied compensation field over the search range while parametrically exciting at some fixed frequency offset to the approximation just determined, and determine the point of minimal excitation by a complex linear fit to the count rate correlations (see~\figref{fig:closeup}).
\end{enumerate}
Steps 1–3 are applied, alternating between all motional modes, until the change in estimate between iterations is smaller than some preset target accuracy.
After having cycled through each mode two times the mode direction estimates can be refined, allowing compensation voltage sets to be chosen adaptively such as to minimally disturb the result of previous iterations.
As the \RF{} null is approached, the excitation amplitude can be increased while decreasing the scan radius to improve the sensitivity.
The mode frequency calibrations, i.e.~steps 1–2, can be skipped if the mode parameters are well-known, for instance when repeatedly running calibrations to track small stray field changes while the setup remains nominally unchanged.

A sample compensation scan is shown in \figref{fig:closeup}.
Matching the prediction from \eqnref{eq:correlated-ampl-phase} and \eqnref{eq:correlated-rate-demod}, the correlated rate is observed to change linearly with the applied electric field.
The setting that minimises the micromotion amplitude is thus easy to determine by linear regression, which is computationally cheap and robust to statistical perturbations.

In this dataset, an effect not captured by the pseudopotential approximation is visible:
Instead of crossing the origin, the line describing the correlated amplitude is offset from zero by a small amount, here corresponding to a correlated fraction of \num{0.006} (a displacement equivalent to a stray field of \SI{0.12}{\volt / \metre} along the fit line).
Numerical simulations of \eqnref{eq:decoupled-mathieu-damped} with an added uniform $\sin(\wRF t)$ term indicate that a possible origin would lie in a residual out-of-phase \RF{} field of $\ish \SI{60}{\volt / \metre}$ near the ion position, for instance caused by slight imbalances in drive or grounding of the two \RF{} electrodes.
Similar to the equivalent effect in \RF{} correlation measurements, for which an analytical approximation is more readily derived \cite{berkelandMinimizationIonMicromotion1998,kellerPreciseDeterminationMicromotion2015}, this appears to shift the observed complex rate in a direction precisely perpendicular to the modulation caused by stray \DC{} fields, thus not affecting the inferred compensation optimum.

The main source of measurement uncertainty is the photon shot noise, i.e. the Poissonian distribution of observed photon counts.
As such, large light collection efficiencies are desirable to reduce data acquisition time for a given target accuracy.
Beyond that, there is a tradeoff between sensitivity and robustness:
For operation in automated calibration routines, we typically choose larger damping rates and detunings ($\ish \SI{10}{\kilo\hertz}$) for robustness against slow radial mode frequency drifts, as well as lower excitation amplitudes to handle larger stray field changes without leaving the linear region.
With these conservative settings, stray fields can be compensated to well below \SI{1}{\volt/\metre} from \num{200000} photons per direction in our apparatus.

When choosing larger excitation amplitudes and smaller detunings ($\ish \SI{3}{\kilo\hertz}$), sensitivities of the correlated fraction to stray fields of $|S| / E \approx \SI{1.4}{\kilo\hertz/ (\volt \metre^{-1})}$ at total count rates of $\eta R_0 = \SI{17}{\kilo\hertz}$ can routinely be reached in our apparatus, giving a statistical noise floor of $\alpha = \SI{0.1}{V m^{-1} / \sqrt{\hertz}}$ per mode direction.
Slow, technical drifts in the voltage sources and the environment (e.g.~due to temperature changes) limit the ultimate uncertainty in our current setup to approximately $\SI{0.015}{\volt / \metre}$, equivalent to a positional uncertainty of $\SI{0.05}{\nano\metre}$ or a force of $\SI{2}{\zepto\newton}$, achieved at an averaging time of $\approx \SI{150}{\second}$ (see appendix~\ref{sec:stability}).

We also implement two phase-modulation-based methods against which to compare the parametric excitation results (albeit only along the one direction parallel to the trap plane in which the relevant laser beams propagate).
Here, we find small, but statistically significant differences between the three approaches:
A typical set of compensation fields obtained by the parametric excitation method is $E_{\mathrm{IP}} = \SI{-119.6(1)}{\volt / \metre}$ and $E_{\mathrm{OOP}} = \SI{147.8(1)}{\volt / \metre}$.
Subsequently changing $E_{\mathrm{IP}}$ to minimise the correlations between \SI{422}{\nano\metre} photons scattered during Doppler cooling and the trap \RF{} drive yields a value lower by \SI{-1.0(1)}{\volt / \metre}.
On the other hand, the minimum in Rabi frequency on the first negative micromotion sideband of the narrow-linewidth $\left|{S}_{1/2}, m=-1/2 \right\rangle \leftrightarrow \left|{D}_{5/2}, m=-3/2 \right\rangle$ transition (\SI{674}{\nano\metre}) occurs at an $E_{\mathrm{IP}}$ value increased by \SI{1.1(1)}{\volt / \metre}.

The origin of these shifts remains unclear.
Radiation pressure from \SI{422}{\nano\metre} photons can be excluded as a reason: the total force is $\ltish \SI{6}{\zepto\newton}$, equivalent to an electric field of $\SI{0.04}{\volt \metre^{-1}}$.
The \RF{} correlation method is, in contrast to the parametric excitation and \SI{674}{\nano\metre} sideband ratio methods, affected by a substantial modulation in PMT detection efficiency (see appendix~\ref{sec:pmt-gain-systematics}), but the above results have been corrected for this.
With the exception of a $\ish \SI{1}{\volt / \metre}$ shift in some cases where the linear approximation breaks down (for strong modulation and positive detunings, which can be avoided by monitoring the total fluorescence; see appendix~\ref{sec:detuning-shift}), no systematic dependence of parametric excitation results on the modulation parameters was observed.
We note that a discrepancy between \RF{} correlation and sideband methods of similar magnitude has been reported in \citeref{kellerPreciseDeterminationMicromotion2015}.

\section{Discussion}

We have described a method for stray field compensation in radio-frequency traps based on parametric excitation of the secular modes of motion and synchronous detection of particle motion.
No directional resolution is required for the motional measurement, as directions are addressed via spectrally separated oscillatory modes.
For trapped ions, this means a single laser cooling beam is sufficient for compensation in all spatial directions as long as it has a projection onto all modes, which is anyway required for efficient laser cooling.
Neither the ability to access transitions of linewidths much smaller than the trap drive frequency nor to perform cooling to the motional ground state is required.
The presented method is thus equally applicable to other types of \RF{}-trapped systems, as long as motion can be weakly damped and detected.

Phase sensitivity in the detection of motion is not strictly necessary; the compensated point could be found by just minimising the magnitude of parametrically excited motion.
When phase information is available, however, the compensation optimum can trivially be extracted in a robust fashion even in the presence of large amounts of statistical noise by linear regression, making this technique well-suited to unattended calibration routines.
Compared to approaches based on variations in the total fluorescence rate due to parametric instabilities, where the change in rate away from the \RF{} null can be positive or negative depending on the parameters, the signal here is unambiguous and does not require special care to develop robust fitting algorithms.
As the method utilises excitation amplitudes below the parametric instability threshold and works in a small-amplitude regime where the cooling dynamics are to good approximation linear, the trap stability is hardly affected; we do not observe elevated ion loss rates.

The parametric excitation method is simple to implement in practice. Technical requirements are limited to a provision for amplitude modulation of the trap potential, whether through a built-in function of the trap \RF{} source or an external circuit.
The required modulation depth is small, so the necessary amplitude-modulation sidebands can be passed through all but the most narrow-band step-up resonator designs.
Note that \eqnref{eq:pseudopotential-damped-mathieu} has the same form as the equations of motion for a static harmonic potential superposed with an oscillating electric field linear in position, $E_x(t) = \wAM^2\, m\, Q\, x\, /\, (2\, \charge) \cos(\wAM t)$.
If amplitude modulation is not feasible, the parametric excitation signal could thus also be applied directly to the \RF{} electrodes, bypassing any resonant circuits using a diplexer~\footnote{Conversely, techniques where a modulating potential is directly applied to the trap \RF{} electrodes, such as the quantum squeezing protocol demonstrated in \citeref{burdQuantumAmplificationMechanical2019}, should be accessible through amplitude-modulation sideband near $\wRF$ as well.}.

Compared to techniques which detect stray fields by imaging the shift in ion position resulting from changes in \RF{} trap strengths, an advantage of this scheme is that the trap configuration can be kept nearly constant, alleviating any concerns about systematic shifts due to changes e.g.~in power dissipation.
The required photon timestamp resolution is set by the motional frequencies; with typical secular mode frequencies in the low \si{\mega\hertz} region, phase jitter small enough not to significantly reduce the signal can easily be achieved using almost all commercially available photon detector and real-time control hardware.

While the method relies on knowledge of well-separated secular mode frequencies, the width of the parametric resonance feature can be tuned via the cooling parameters.
We use this scheme as the primary method for micromotion minimisation in a system designed without any focus on radial mode stability or active stabilisation of the trap \RF{} amplitude, where mode frequency changes of several kilohertz due to laboratory temperature fluctuations are observed.

\newcommand{\tabcell}[2][c]{\begin{tabular}[#1]{@{}c@{}}#2\end{tabular}}
\begin{table}
	\begin{tabular}{c c c}
		& \tabcell{$\sigma_E$ \\ \footnotesize{$\si{V m^{-1}}$}} & \tabcell{$\alpha$ \\ \footnotesize{$\si{V m^{-1} / \sqrt{Hz}}$}}\\
		\hline
		Cavity emission spec.~\cite{chuahDetectionIonMicromotion2013} & 1.8 & 1.6 \\
		Repumper \RF{} correlations~\cite{allcockImplementationSymmetricSurfaceelectrode2010} & 1 & 1 \\
		Micromotion sideband Rabi freq.~\cite{chwallaPrecisionSpectroscopy40Ca2009} & 0.4 & - \\
		Param.~excitation (total fluo.) \cite{kellerPreciseDeterminationMicromotion2015} & 0.3 & - \\
		Focus-scanning imaging \cite{glogerIontrajectoryAnalysisMicromotion2015} & 0.09 &	2.30 \\
		\RF{} correlations \cite{kellerPreciseDeterminationMicromotion2015} & 0.09 & 0.57 \\
		Co-trapped ultracold atoms \cite{harterMinimizationIonMicromotion2013} & 0.02
		& - \\
		This work & 0.015 & 0.1 \\
	\end{tabular}
	\caption{
		Various experimental stray field compensation results.
		For each method, the lowest published stray field uncertainty $\sigma_E$ given, along with the inverse sensitivity $\alpha$ where reported for the same measurement.
		The scaling to other quantities of interest, e.g.~the induced second-order Doppler shift, varyingly depends on ion species and trap parameters; see \citeref{kellerPreciseDeterminationMicromotion2015} for a more detailed comparison.
	}
	\label{tab:literature}
\end{table}

The achieved statistical uncertainties are comparable to or below that of the lowest results reported in the literature for other stray field compensation methods (see table~\ref{tab:literature}).
In addition to finding the stray field minimum, this method can in principle also be used to directly obtain single-point estimates of the \RF{} potential at the ion position owing to the linear response.
For this application, however, either a detailed description of the ion cooling dynamics or the capability to apply well-characterised additional fields for calibration (i.e.~a locally accurate model for the \DC{} electrode fields) is necessary to calculate the scaling factor.

While numerical results indicate a small sensitivity to out-of-phase \RF{} field terms such as caused by phase asymmetries in the \RF{} electrode drive, this can not be described in the simple pseudo-potential picture presented here. This method is hence not particularly suitable for the investigation of such issues \cite{berkelandMinimizationIonMicromotion1998,mohammadiMinimizingRfinducedExcess2019}.

In this manuscript, we have only discussed a mode of operation where both cooling and parametric excitation close to the motional frequency ($\wAM \approx \wm$) are applied in a stationary fashion, as the resulting signal is robust, linear in the stray field, and independent of initial conditions.
To increase sensitivity if the available modulation depth is limited, this scheme could be combined with excitation at the principal parametric resonance ($\wAM \approx 2 \wm$) to amplify small amounts of motion \cite{yuNoiseFreeParametric1993}.
By preparing the motional modes in a sufficiently well-known state through initial coherent excitation (possibly after ground state cooling), parametric excitation could also be used to later probe stray fields where no dissipation mechanism is available, for instance along laser-free storage or interaction regions of an extended shuttling-based quantum computing chip.
It should also be possible to acquire data for multiple mode directions in parallel by applying multiple amplitude-modulation tones at the same time.

In conclusion, we have presented a simple method for multi-directional stray field compensation in Paul traps based on spectrally addressed parametric excitation of the secular modes through amplitude modulation of the \RF{} trapping potential.
The technique is applicable to a wide variety of trapped particle systems; in typical trapped-ion experiments, the added technical complexity is minimal, and we have demonstrated resolution competitive with that reported for alternative methods.

\section*{Acknowledgements}

We would like to thank T.~G.~Ballance for initially suggesting this line of inquiry, and J.~F.~Goodwin for helpful comments on the manuscript.
B.~C.~N.\, acknowledges funding from the U.K.~National Physical Laboratory.
C.~J.~B.\, acknowledges support from a UKRI FL Fellowship, and is a Director of Oxford Ionics Ltd.
This work was supported by the U.K.~EPSRC \enquote{Networked Quantum Information Technology} Hub and the EU~Quantum Technology Flagship Project AQTION (No.\ 820495).

\clearpage

\appendix

\section{Approximate solutions of the damped, inhomogeneous Mathieu equation}
\label{sec:analytical-mathieu-solns}

Consider the damped, inhomogeneous Mathieu-type equation
\begin{equation}
	\difftwo{x}{T} + 2 Z \diff{x}{T} + \left(A + 2 Q \cos 2 T + R \cos 4 T\right) x = F,
	\label{eq:normalised-mathieu}
\end{equation}
where the extra term at twice the frequency is motivated by the pseudopotential approximation from \eqnref{eq:pseudopotential-damped}, and we have transitioned to normalised time by substituting $T \defeq \frac{\wAM}{2} t$ and writing $Z \defeq \frac{2 \zeta}{\wAM}$, $A \defeq \left(\frac{2 \wm}{\wAM}\right)^2$, $Q \defeq \hred A$, $R \defeq h \hred A$, $F \defeq \left(\frac{2}{\wAM}\right)^2 \frac{E \charge}{m}$.

Here, we are interested in bounded solutions for $A \approx 4$, corresponding to cases where $\wAM \approx \wm$.
All coefficients in \eqnref{eq:normalised-mathieu} are periodic in time with period $\pi$, so informed by a Floquet analysis of the system, we search for solutions of the form
\begin{equation}
	x(T) =\sum_{n = -\infty}^\infty c_n \ee^{2 \ii n T},
	\label{eq:harmonic-ansatz}
\end{equation}
with coefficients $(c_n)_{n \in \mathbb{Z}} \subset \mathbb{C}$ satisfying $c_{-n} = \left(c_n\right)^{*}$ as $x(T) \in \mathbb{R}$.
Inserting \eqnref{eq:harmonic-ansatz} into \eqnref{eq:normalised-mathieu}, i.e.~following a harmonic balance approach, yields a system of equations for the coefficients
\begin{multline*}
	2 Q (c_{n-1} + c_{n+1}) + R (c_{n-2}+c_{n+2}) \\ + 2(A - 4 n^2 + 4 n \ii Z) c_n = \delta_{n,0} 2 F \ .
\end{multline*}
The magnitude of $c_n$ will fall rapidly with $|n|$ due to the $4 n^2$ factor, as $Q$, $R$, and $Z$ are small for typical conditions.
We can thus obtain an analytical solution by truncating the system at an appropriate $|n|$ for the desired degree of accuracy.

To do so for small excitation amplitudes, we substitute $Q = \hred A$ and $R = \alpha \hred^2 A$ (with $\alpha \defeq h / \hred \approx 1$), and expand to second order in $\hred$ (for which including terms up to $|n| = 2$ is sufficient), obtaining
\begin{equation}
	\begin{gathered}
		c_0 = \frac{F}{A}\left(1 - \hred^2 \frac{2 A (4 - A)}{(4 - A)^2 + (4 Z)^2}\right) + O(\hred^3)\ , \\
		c_1 = \hred \frac{F}{4 - A - 4 \ii Z} + O(\hred^3) \ , \\
		c_2 = \hred^2 \frac{F(2A + \alpha(4 - A - 4 \ii Z))}{2(4 - A - 4 \ii Z)(16-A-8\ii Z)} + O(\hred^3) \ .
	\end{gathered}
	\label{eq:second-order-solution}
\end{equation}

It is worth noting that the ansatz (\ref{eq:harmonic-ansatz}) only covers bounded solutions, and only that of excitation near integer subharmonics of the motional frequency (of which the case $\wAM \approx \wm$ is the strongest).
Parametric resonance near $2 \wm$ and other subharmonics thereof excites motion at half the modulation frequency, which is consequently not captured by the above solution.

Analytical approximations for the stability boundaries can be obtained for small damping factors and detunings from resonance using a multiple-parameter perturbation method \cite{wuPerturbationMethodFloquet1995}.
We have verified that $\hred = \frac{2 \zeta}{\wm}$ is an excellent approximation for the stability boundary for zero detuning in the experimentally relevant regime by numerical integration of \eqnref{eq:decoupled-mathieu-damped}, although in practice nonlinearities in laser cooling and trap potentials significantly affect the transitory regime towards instability.

\section{RF generation setup}
\label{sec:rf-setup}

\begin{figure}
	\includegraphics[width=0.8\linewidth]{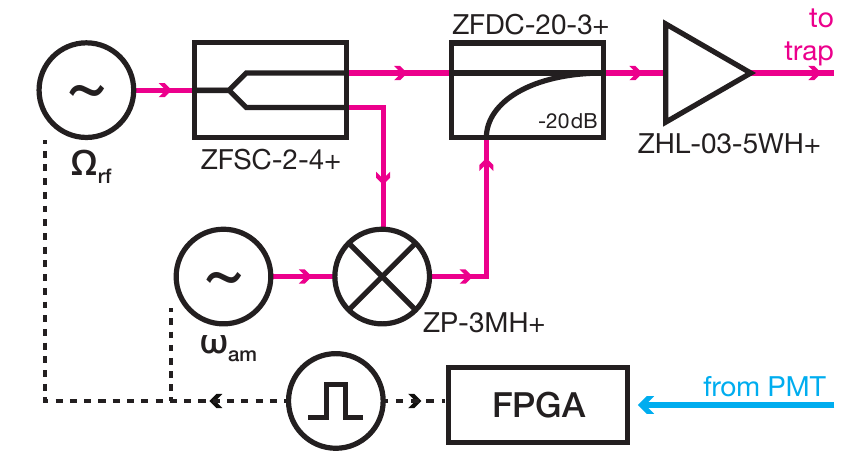}
	\caption{
		Block diagram of the \RF{} generation setup employed for this demonstration (part numbers of suitable MiniCircuits components given for reference).
		Both the Urukul DDS \RF{} sources and the \FPGA{} use a common reference clock, making it trivial to accurately demodulate photon timestamps in the digital domain.
	}
	\label{fig:rf-chain}
\end{figure}

To generate the required amplitude-modulated trap \RF{} drive, we used two channels of a \emph{Urukul} direct digital synthesis card from the ARTIQ/Sinara open-hardware ecosystem \cite{ARTIQ}.
As shown in \figref{fig:rf-chain}, one channel at $\wRF$ supplied the unmodulated trapping potential (though an additional limiting amplifier for some of the data).
To provide a small, tunable amount of amplitude modulation without affecting the mode stability in regular operation, the signal was split into two paths using a passive power splitter (MiniCircuits ZFSC-2-4-S+), one of which was modulated using a double-balanced diode-ring mixer (MiniCircuits ZP-3MH-S+), whose IF port was driven from the second Urukul channel set to $\wAM$.
The signals were then recombined using a directional coupler (MiniCircuits ZFDC-20-3-S+) and amplified (MiniCircuits ZHL-03-5WH+) before the single-stage $LC$ trap resonator.

In this configuration, the phase of both the $\wRF$ and $\wAM$ tones is stable and known in absolute terms with respect to the \FPGA{} timeline, enabling simple demodulation of the photon arrival times in the digital domain for both the parametric excitation and \RF{} correlation methods, without the need to read back the analogue signal again.
Note that while this approach is both highly accurate and convenient to implement on our experimental system, many other demodulation strategies are possible, as the timing precision and stability requirements for the parametric excitation method are relatively modest.
Even in case the source used to generate the amplitude modulation signal cannot directly be phase-synchronised with the device used to timestamp the photon clicks, the $\wAM$ tone could simply be time-tagged along with the signal using an additional channel.
The relevant time scale for the demodulation of the photon arrival times is given by the inverse of the modulation frequency $\wAM \approx \wm$.
For typical  $\ish \si{\mega\hertz}$-level mode frequencies, even a microprocessor-based system or a \FPGA{} platform running at moderate clock frequencies and without any high-resolution input timestamping techniques would thus provide sufficient resolution.

To adjust the applied modulation index, the output amplitude and step attenuator setting of the $\wAM$ tone are adjusted.
The actual modulation index of the trapping potential is reduced compared to the generated RF signal by the band-pass response of the resonant impedance matching circuit.
The values for $\h$ quoted throughout this manuscript were obtained by sampling the signal at a capacitive divider after the resonator using an oscilloscope and computing the ratio of modulation sideband to carrier powers in the resulting Fourier spectrum.
As this is dependent on the exact response of the resonator, which drifts slowly over time, the quoted values for $h$ should not be taken to be more accurate than $\ish\SI{10}{\percent}$.

Note that in the above configuration, any phase difference between the direct and modulated paths will lead to a more complex modulation response in place of pure amplitude modulation.
For the purposes of this demonstration, the two arms were matched to within $\SI{25(3)}{\pico\second}$, or, equivalently, $\SI{8(1)}{\milli\radian}$ at $\wRF \approx \SI{50}{\mega\hertz}$.
However, numerical simulations indicate that the presence of even a large phase-modulation-like component is not detrimental to the accuracy of the compensation method.

\section{Mode frequency shift with excitation amplitude}
\label{sec:mode-freq-shift}

\begin{figure}
	\includegraphics{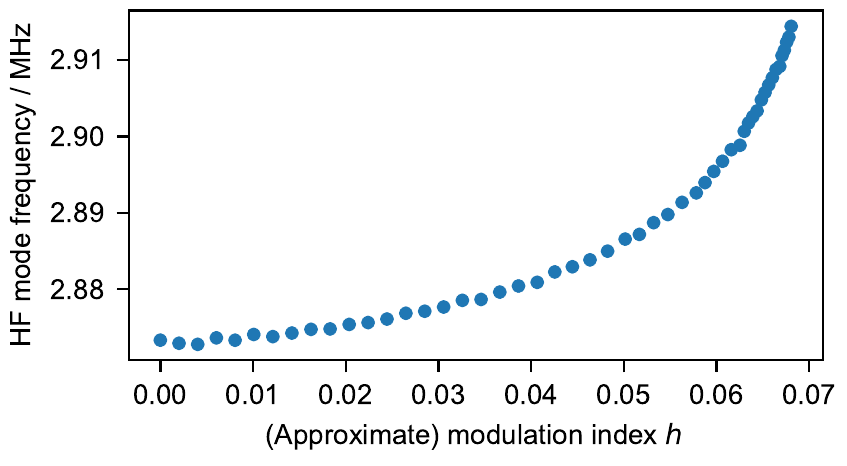}
	\caption{Shift of the HF mode frequency versus amplitude of an off-resonant amplitude modulation tone at $\wAM = 2\pi \times \SI{2.5}{\mega\hertz}$, as observed using a \enquote{tickle} measurement. The statistical uncertainty of each point is smaller than the marker size, but the accuracy is affected by \si{\kilo\hertz}-level drifts over the duration of the measurement (points acquired in random order).}
	\label{fig:mode-freq-shift}
\end{figure}

In the pseudopotential approximation, a slight increase in mode frequency with increasing modulation index $h$ is expected according to \eqnref{eq:pseudopot-modefreq}.
On the other hand, numerical integration of \eqnref{eq:decoupled-mathieu-damped}, the full equation of motion, for parameters $a = -0.0026$ and $q = 0.17$ suggests a decrease in observed mode frequency for pure amplitude modulation instead.

To probe this effect in isolation, trap \RF{} modulation was applied at $\wAM = 2\pi \times \SI{2.5}{\mega\hertz}$, which is far from resonance with either motional mode.
The HF mode frequency was then determined using the \enquote{tickle} method, where a resonant electric field applied to one of the trap electrodes excites the ion motion. We detect the change in total fluorescence rate due to the large motional amplitudes produced this way. An increase of mode frequency with modulation index is indeed observed, but unfortunately, the exact response is not modelled well by either approach – if it were, this could serve as a convenient calibration of the actual modulation index after the band-pass filter formed by the resonant matching circuit.

In practice, this shift is not a hindrance for use of the micromotion compensation method, as the mode frequency can simply be determined at the target modulation index.
However, it needs to be accounted for when employing a different method to calibrate the mode frequencies used to derive the amplitude-modulation settings and large modulation depths are to be used.

\section{Statistical error and stability}
\label{sec:stability}

\begin{figure}
	\includegraphics{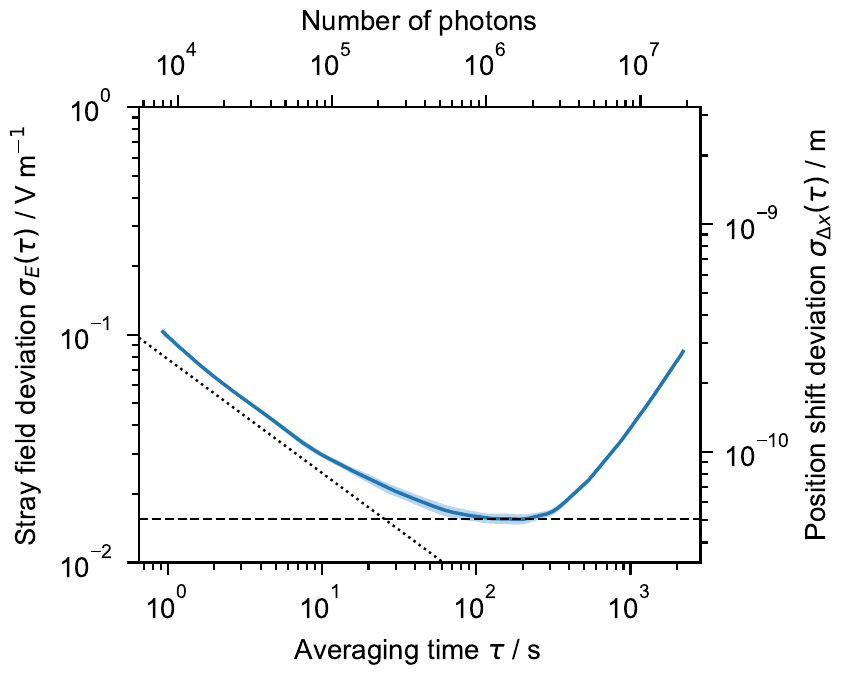}
	\caption{
		Stability evaluation of the compensation field in the direction of the high-frequency mode. Shown is the overlapping two-sample (Allan) deviation $\sigma_E(\tau)$ for the inferred compensation field for different averaging times $\tau$, converted into a positional uncertainty for the $\SI{2.92}{\mega\hertz}$ \srplus{} mode on the right axis. The shaded region indicates the statistical uncertainty ($1 \sigma$) due to photon shot noise, obtained from parametric bootstrapping. The dotted line denotes the error expected purely from photon shot noise at the fitted sensitivity, and the dashed line indicates the minimum uncertainty of $\sigma_E = \SI{0.015}{\volt \metre^{-1}}$ achieved at an averaging time of $\tau \approx \SI{150}{\second}$.}
	\label{fig:comp-variance}
\end{figure}

To estimate the resolution limit of our method given the slow drifts affecting various experimental parameters, we consider Allan's two-sample variance for the inferred stray electric field, as shown in \figref{fig:comp-variance}.
We continuously measure the correlated amplitude for a range of applied compensation fields chosen uniformly between $\pm\SI{2.5}{\volt \metre^{-1}}$, for excitation with $h \approx 0.03$ detuned $\SI{-3}{\kilo\hertz}$ from the high-frequency radial mode, cooling parameters $s \approx 1$, $\Delta = \SI{-13}{\mega\hertz}$, and $N = \num{1000}$ photons per compensation field setting.

Let $\bar{E}_{i;l}$ denote the estimate for the compensation field that minimizes the excess micromotion amplitude as determined by linear regression (as discussed in \secref{sec:compensation}; see \figref{fig:closeup}) over the subset of correlated amplitude measurements from index $i$ to $i + l - 1$, which we assume to originate from $L - 1$ equal time intervals $\Delta t = \frac{N}{\eta R_0}$.
For technical simplicity, we acquire an equal number of photons per point, making this only true in approximation due to the Poissonian count statistics.
The overlapping two-sample variance ${\sigma_E(\tau)}^2$ for an averaging period $\tau = l\, \Delta t$, $l \in \mathbb{N}$, is then
\begin{equation}
	\sigma_E^2(l\, \Delta t) = \frac{1}{2 (L - 2 l + 1)} \sum_{i = 1}^{L - 2l} \left(\bar{E}_{i; l} - \bar{E}_{i + l; l}\right)^2.
\end{equation}

Figure~\ref{fig:comp-variance} shows the dependence of variance on the averaging time for a data set typical for our experimental setup, for averaging windows of $l = 8$ to $\num{19640}$.
At short times, the variance drops with the square root of the averaging time, reflecting the floor set by the photon shot noise.
Around $\tau \approx \SI{1}{\minute}$, a plateau is reached, where further averaging does not significantly improve the variance.
The achieved minimal uncertainty is  $\sigma_E = \SI{0.015}{\volt \metre^{-1}}$, which, given the \SI{2.92}{\mega\hertz} frequency of the mode used for this dataset and the charge-to-mass ratio of \srplus{}, corresponds to an uncertainty in position relative to the \RF{} null of  $\sigma_{\Delta x} = \SI{0.05}{\nano\metre}$, or an equivalent force of $\sigma_F = \SI{2}{\zepto\newton}$.
At longer averaging times, slow drifts in the stray fields or electrode voltages dominate.

\section{RF modulation of detector efficiency}
\label{sec:pmt-gain-systematics}

As previously reported by Keller et al.~\cite{kellerPreciseDeterminationMicromotion2015}, we find that the quantum efficiency of the photomultipler tube used to detect the ion fluorescence is modulated by the presence of ambient radio-frequency fields, in particular by radiation leaking from the trap electrodes.
Extraneous frequency content near the carrier frequency has the potential to cause systematic shifts in any measurement relying on synchronous demodulation, such as the \RF{} correlation and parametric excitation micromotion compensation schemes:

If the detection probability of photons is modulated by $G(t) = 1 + \re{g\, \ee^{\ii \wRF t}}$, where $g$ is the complex modulation amplitude, a signal of amplitude $r \in \mathbb{C}$ at frequency $\wRF$ in the fluorescence rate  $R(t) = R_0 (1 + \re{r \, \ee^{\ii \wRF t}})$ will yield a demodulated amplitude of  $\langle G(t) R(t) \ee^{-\ii \wRF t}\rangle_t \propto r + g$ instead, i.e.~the efficiency modulation results in a shift of the measured signal in the complex plane.
In the context of stray field compensation, a component of $g$ along $r$ will cause a systematic offset in the inferred minimum.

The phase difference between $r$ and $g$ depends on various phase shifts and time delays, whether in the \RF{} signal chain (e.g.~the physical detector placement) or the ion phase response (influenced by the cooling laser parameters).
Thus, to obtain accurate results using the \RF{} correlation method, the gain modulation must be characterised so it can be subtracted from the data.

For the data presented here, we use a Hamamatsu H10682-210 photomultiplier module to detect ion fluorescence, situated approximately \SI{0.5}{\metre} from the trap centre.
With the ion trap empty, light from a torch is leaked into the apparatus to provide photon counts at a rate close to that used for the \RF{} correlation measurements without any possibility for correlations with the trap \RF{} drive.
A modulation with a magnitude of \SI{1.2}{\percent} is measured and subsequently subtracted from the \RF{} correlation data, leading to a shift of \SI{2.3}{\volt \metre^{-1}} in the micromotion minimum estimated using that method.
As discussed in \secref{sec:compensation}, a small difference between the results of the \RF{} correlation and parametric excitation methods remains even taking this offset into account.

While the exact origin of this effect is not known, there is good reason to expect the parametric excitation method proposed here to be far less susceptible to it.
Not only is the \RF{} amplitude only modulated by a small factor $h \ll \SI{1}{\percent}$, if the detection chain responds linearly, the signal will only exhibit frequency components at $\wRF \pm \wAM$ (but not at $\wAM$).
Even for a large modulation depth $h = \SI{3}{\percent}$, the correlated amplitude for background light was measured to be consistent with zero, $S =\num{6(11)e-5}$.

\section{Frequency-dependent shift for large modulation amplitudes}
\label{sec:detuning-shift}

\begin{figure}
	\includegraphics{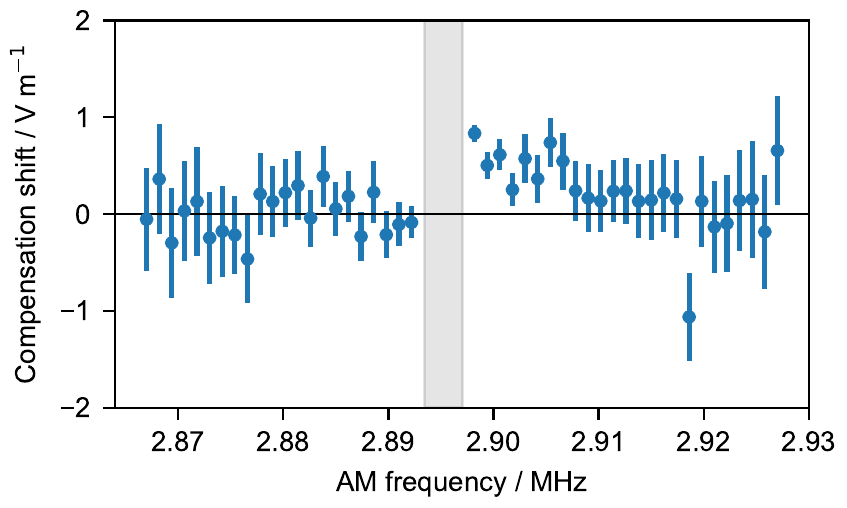}
	\caption{Systematic shift in the inferred compensation optimum at large excitation amplitudes: Each point represents the compensation optimum inferred from a fit to \num{51} correlation measurements at offsets of $\pm \SI{10}{\volt \metre^{-1}}$ along the HF mode direction, with modulation index $\h \approx 0.02$ and cooling parameters $\Delta = -2\pi \times \SI{13}{\mega\hertz}$, $s \approx 0.9$. Immediately above the mode frequency, indicated by the shaded region (including measurement uncertainty and drifts over the measurement duration), a systematic shift towards positive values is apparent.}
	\label{fig:detuning-shift}
\end{figure}

According to the theory developed in \secref{sec:theory}, the stray field compensation optimum is unambiguously given by the point where the correlated amplitude vanishes.
The modulation frequency only affects the strength (and, close to resonance, phase) of the correlation signal.

At low and moderate modulation depths $\h$, the correlation optimum, as inferred by the complex linear regression procedure, is indeed observed to be independent of the choice of modulation frequency $\wAM$, enabling robust stray field compensation. At high modulation depths, however, we observe a systematic shift of up to \SI{1}{\volt \metre^{-1}} (compared to the value obtained at lower excitation strengths) for $\wAM$ slightly above the mode frequency.
An example of this behaviour is shown in \figref{fig:detuning-shift}.

The exact mechanism for this is unknown and the behaviour could not be replicated in numerical simulations using \eqnref{eq:decoupled-mathieu-damped}.
Empirically, however, this shift is only observed if the ion is driven strongly enough to where the response is no longer linear and, for $\wAM$ close to the mode frequency, a parametric instability is excited.
This can be diagnosed by its effect on the total fluorescence rate, which remains constant in the linear regime.
By choosing the experimental parameters such that the fluorescence does not significantly change as $\wAM$ is scanned over the resonance, or as the modulation is turned on and off at the target $\wAM$, the problematic parameter regime can easily be avoided in practice.

\bibliography{parametric-compensation-servo}

\end{document}